\documentclass[intlimits,twoside,a4paper]{article}

\usepackage{amsmath,amssymb}
\usepackage{graphicx}
\usepackage{wrapfig}

\usepackage[T2A]{fontenc}
\usepackage[cp1251]{inputenc}

\usepackage{epsf,multicol,ifthen}

\usepackage[eqsecnum]{cmpj3}

\issue{2019}{22}{1}{13801}
\doinumber{10.5488/CMP.22.13801}

\title[The effect of the electric field on the nucleation of the nanometer periodic structure of adatoms in GaAs ]%
{The effect of the electric field on the nucleation of the nanometer periodic structure of adatoms in GaAs semiconductor under the action of laser irradiation}
\author[R.M.~Peleshchak, O.V.~Kuzyk, O.O.~Dan'kiv,  S.K.~Guba]{R.M.~Peleshchak\refaddr{label1}, O.V.~Kuzyk\refaddr{label1}, O.O.~Dan'kiv\refaddr{label1}\thanks{E-mail: dankivolesya@ukr.net}\,,  S.K.~Guba\refaddr{label2}}
\addresses{
\addr{label1} Drohobych Ivan Franko State Pedagogical University, 24 Franko St., 82100 Drohobych, Ukraine 
\addr{label2} Lviv National Polytechnic University, 12 Bandera St.,  79013 Lviv,  Ukraine
}

\date{Received August 27, 2018, in final form December 9, 2018}

\begin{document}

\maketitle

\begin{abstract}
In the paper, the effect of the electric field on the conditions of formation and on the period of the surface superlattice of adatoms in \textit{n}-GaAs semiconductor is investigated. It is established that in GaAs semiconductor, an increase in the electric field strength, depending on the direction, leads to an increase or decrease of the critical temperature (the critical concentration of adatoms), at which the formation of self-organized nanostructure is possible. It is shown that in strongly alloyed \textit{n}-GaAs semiconductor, an increase of the electric field strength leads to a monotonous change (decrease or increase depending on the direction of the electric field) of the period of self-organized surface nanostructures of adatoms.
\keywords nucleation, electric field, adatom, surface superlattice, diffusion, deformation
\pacs 81.07.Bc, 66.30.Lw
\end{abstract}

\section{Introduction}

Recently, the possibility of obtaining semiconductor structures with self-organized nanoclusters using methods of molecular-beam epitaxy~\cite{Zin15,Ipa98}, ion implantation~\cite{Yam03,Bor12} and under the action of laser irradiation~\cite{Pel10,Vla15}, as well as the ability to control their physical properties, have become the subject of intense research. In particular, laser-induced periodic surface nanostructures can be generated practically on  any material (metals, semiconductors, dielectrics) at linearly polarized irradiation and are formed in a wide range of intervals of impulses, ranging from continuous wave radiation to several femtoseconds~\cite{Hoh13,Wu17,Bon17,Eme08}. In the experimental work~\cite{Vla15} it was shown that their formation is caused by the effect of a long-range action of a laser pulse and is explained by the effect of the pressure gradient of the surface acoustic wave. The data on nucleation (incipient state of the formation) of periodic nanostructures of adsorbed atoms (adatoms) and implanted impurities are important for optimization of the technological process and for a predictable control of the physical parameters of semiconductor structures with nanoclusters. In particular, in order to calculate the symmetry, the period and the formation time of surface structures, it is sufficient to analyze only the initial (linear) stage of the development of the defect-deformation instability.

In~\cite{Eme08}, there was developed a theory of spontaneous nucleation of the surface nanometer lattice which is due to instability in the system of adatoms interacting with self-consisting surface acoustic wave (SAW). Within this theory, the conditions of formation of nanoclusters on the surface of solids and the periods of nanometer lattice as a function of concentration of adatoms and temperature are defined.

In the experimental works~\cite{Zen16,Tan17} the effect of the external electric field on the formation of self-organized nanostructures was investigated. In particular, in~\cite{Tan17} it is shown that the external electric field increases the density and changes the size of the CuO quantum wires.

The elastic fields created by defects turn out to be a determining factor in the formation of the surface superlattice of adatoms. In works~\cite{Pel18,Tay08,Pelu16,Pel016,Pel13} it is shown that external hydrostatic pressure, ultrasonic wave and doping with isovalent impurities should contribute to the improvement of the conditions for the formation of surface nanostructures. The periodic deformation arising on the surface of a semiconductor leads to the modulation of the bottom of the conduction band and, consequently, to the modulation of electronic density. The arising nonuniform electric field leads to a nonuniform displacement of the nodes of the crystal lattice and, consequently, to the change in the amplitude of the SAW~\cite{Pel16,Pel15}. Therefore, it can be expected that when placing a semiconductor in the external electric field, it is possible to change the conditions of the formation of laser-induced periodic surface nanostructures and predictably control their parameters due to the interaction of the electric field with nonuniformly distributed free current carriers on the surface.

Due to the high mobility of charge carriers, gallium arsenide is widely used in the production of quantum-sized structures and high-frequency lasers on their basis. Considerable attention of researchers is paid to laser modification of the morphology of near-surface layers of GaAs, in which the conditions of the formation of nanoclusters are a controlled process~\cite{Wu17,Yaz11}. In addition, gallium arsenide has one of the largest constants of the deformation potential of the conduction band (greater only in Ge)~\cite{Chr89}, which makes the conduction band and the electronic subsystem sensitive to deformation. Therefore, it is expected that the interaction of the electric field with periodic deformation will be quite significant. In this paper, the effect of the external electric field perpendicular to the SAW on the nucleation of the nanometer periodic structure of adatoms in GaAs semiconductor is investigated.

\section{The model}

The equations for the displacement vectors $\vec u$ of an elastic medium are of the form~\cite{Lan70}:
\begin{equation}
\label{eq1}
\frac{{\partial ^2 \vec u}}{{\partial t^2 }} = c_\text{t}^2 \Delta \vec u + (c_\text{l}^2  - c_\text{t}^2 )~\textrm{grad}\left( {\textrm{div}~\vec u} \right),
\end{equation}
where $c_\text{l}$ and $c_\text{t}$ are the longitudinal and transversal sound velocities, respectively.

Let the surface of semiconductor coincide with the plane $z = 0$ ($z$-axis is directed into the crystal depth), and assume that along the $x$-axis there is a surface perturbation of the elastic medium, which is given in the form of a static SAW which quickly fades into the depth of a semiconductor and has an amplitude  rowing with time~\cite{Eme08}:
\begin{equation}
\label{eq2}
u_x  =  - \text{i} qR\text{e}^{\text{i}qx + \lambda t - k_\text{l} z}  - \text{i}k_\text{t} Q\text{e}^{\text{i}qx + \lambda t - k_\text{t} z},
\end{equation}
\begin{equation}
\label{eq3}
u_z  = k_\text{l} R\text{e}^{\text{i}qx + \lambda t - k_\text{l} z}  + qQ\text{e}^{\text{i}qx + \lambda t - k_\text{t} z},
\end{equation}
where  $k_\text{l,t}^2  = q^2  + \frac{{\lambda ^2 }}{{c_\text{l,t}^2 }}$;  $\lambda$ is the increment of defect-deformation instability; $R$ and $Q$ are SAW amplitudes.

Then, deformation $\varepsilon$   on the semiconductor surface ($z = 0$) is of the form
\begin{equation}
\label{eq4}
\varepsilon  = \frac{{\partial u_x }}{{\partial x}} + \frac{{\partial u_z }}{{\partial z}} =  - \frac{{\lambda ^2 }}{{c_\text{l}^2 }}R\text{e}^{\text{i}qx + \lambda t}.
\end{equation}

Consider the case where the semiconductor contains impurities --- ionized donors, free electrons and neutral adatoms. The electroneutrality condition will look as follows:
\begin{equation}
\label{eq5}
n_0 = N_\text{d}^ +  ,
\end{equation}
where $N_\text{d}^ +$ and $n_0$ are the surface concentration of ionized donors and the spatially homogeneous values of the surface concentration of electrons, respectively.

Periodic surface deformation leads to  spatial nonuniform redistribution of adatoms $N(x)$, the mo\-du\-la\-tion of the bottom of the conduction band and, respectively, to redistribution of the concentration of conduction electrons $n(x)$ and the electrostatic potential  $\varphi(x)$:
\begin{equation}
\label{eq6}
N(x) = N_0  + N_1 (x) = N_0  + N_1 (q)\text{e}^{\text{i}qx + \lambda t},
\end{equation}
\begin{equation}
\label{eq7}
n(x) = n_0  + n_1 (x) = n_0  + n_1 (q)\text{e}^{\text{i}qx + \lambda t},
\end{equation}
\begin{equation}
\label{eq8}
\varphi (x) = \varphi (q)\text{e}^{\text{i}qx + \lambda t},
\end{equation}
where $N_1(q)$, $n_1(q)$ and  $\varphi(q)$ are the amplitudes of the corresponding periodic perturbations; $N_0$ is the spatially homogeneous values of the surface concentration of adatoms.

The Poisson equation, taking into account (\ref{eq5}), (\ref{eq7}) and (\ref{eq8}), will take the form:
\begin{equation}
\label{eq9}
 - q^2 \varphi (q) = \frac{e}{{\varepsilon _0 \tilde \varepsilon a}}n_1(q)\,,
\end{equation}
where  $\varepsilon_0$ and  $\tilde \varepsilon$ are dielectric constant and dielectric permittivity of the medium, respectively;  $a$ is a lattice constant.

The equations for concentration of adatoms can be presented as follows:
\begin{equation}
\label{eq10}
\frac{{\partial N}}{{\partial t}} = D_\text{d} \frac{{\partial ^2 N}}{{\partial x^2 }} - D_\text{d} \frac{{\theta_\text{d} }}{{k_\text{B} T}}\frac{\partial }{{\partial x}}\left[ {N\frac{\partial }{{\partial x}}\left( {\varepsilon  + l_\text{d}^2 \frac{{\partial ^2 \varepsilon }}{{\partial x^2 }}} \right)} \right],
\end{equation}
where $D_\text{d}$ is the surface diffusion coefficient; $k_\text{B}$ is Boltzmann constant; $T$ is temperature;  $\theta_\text{d}$ is the deformation potential; $l_\text{d}$ is the characteristic length of interaction of adatoms with lattice atoms. The second term expresses the interaction of adatoms with the deformation field, taking into account the nonlocal interaction~\cite{Eme08}. The defect which enters the surface of semiconductor leads to a change in its volume and energy, and the initial fluctuation of deformation under certain conditions causes the emergence of deformation-induced flows of adatoms. In nonuniform deformation-concentration field there are forces proportional to gradients of concentration and deformation.

Taking into account (\ref{eq4}), (\ref{eq6}) and in the approximation of $N_1\ll N_0$, the equation (\ref{eq10}) is written in the form:
\begin{equation}
\label{eq11}
\lambda N_1 (q) =  - D_\text{d} q^2 N_1 (q) - \frac{{D_\text{d} N_0 \theta_\text{d} }}{{k_\text{B} T}}\left[ {\frac{{\lambda^2 }}{{c_\text{l}^2 }}Rq^2 \big(1 - q^2 l_\text{d}^2 \big)} \right].
\end{equation}

The density of the electron current:
\begin{equation}
\label{eq12}
j =- j_\text{el}= n\mu _n \frac{{d\chi }}{{dx}}\,,
\end{equation}
where $ j_\text{el}$ is the flow of electrons; $\mu_n$ is the mobility of electrons; the electrochemical potential $\chi$  is defined by  the relation
\begin{equation}
\label{eq13}
\chi (x) = k_\text{B} T\ln \frac{{n(x)}}{{N_i }} - e\varphi (x) + a_\text{c} \varepsilon (x)\,,
\end{equation}
where $N_i$  is the effective density of states;  $N_i  = 2\left( {\frac{{2\piup mk_\text{B}T}}{{h^2 }}} \right)^{3/2}$; $a_\text{c}$ is the constant of hydrostatic deformation potential of the conduction band. Then, taking into account (\ref{eq12}), (\ref{eq13}), the continuity equation can be written in the form:
\begin{equation}
\label{eq14}
e\frac{{\partial n}}{{\partial t}} = k_\text{B} T\mu _n \frac{\partial }{{\partial x}}\left( {n\frac{\partial }{{\partial x}}\ln \frac{n}{{N_i }}} \right) - e\mu _n \frac{\partial }{{\partial x}}\left( {n\frac{{\partial \varphi }}{{\partial x}}} \right) + a_\text{c} \mu _n \frac{\partial }{{\partial x}}\left( {n\frac{{\partial \varepsilon }}{{\partial x}}} \right).
\end{equation}

Taking into account (\ref{eq4}), (\ref{eq7}) -- (\ref{eq9}), the equation (\ref{eq14}) can be presented as follows:
\begin{equation}
\label{eq15}
n_1 (q)\left( {e\lambda  + k_\text{B} T\mu _n q^2  + \frac{{e^2 n_0 \mu _n }}{{\varepsilon _0 \tilde \varepsilon a}}} \right) = a_\text{c} n_0 \mu _n q^2 \frac{{\lambda ^2 }}{{c_\text{l}^2 }}R.
\end{equation}

From (\ref{eq11}) and (\ref{eq15}), we obtain expressions for the amplitudes of the surface concentration of adatoms $N_1(q)$ and conduction electrons $n_1(q)$:
\begin{equation}
\label{eq16}
N_1 (q) =  - \frac{{D_\text{d} N_0 \theta_\text{d} }}{{k_\text{B} T\left( {\lambda  + D_\text{d} q^2 } \right)}}\left[ {\frac{{\lambda ^2 }}{{c_\text{l}^2 }}Rq^2 \big(1 - q^2 l_\text{d}^2 \big)} \right],
\end{equation}
\begin{equation}
\label{eq17}
n_1 (q) = \frac{{a_\text{c} n_0 \mu _n q^2 }}{{e\lambda  + k_\text{B} T\mu_n q^2  + \frac{{e^2 n_0 \mu_n }}{{\varepsilon_0 \tilde \varepsilon a}}}}\frac{{\lambda ^2 }}{{c_\text{l}^2 }}R.
\end{equation}

\begin{figure}[!b]
	\begin{multicols}{3}
		\includegraphics[width=50mm]{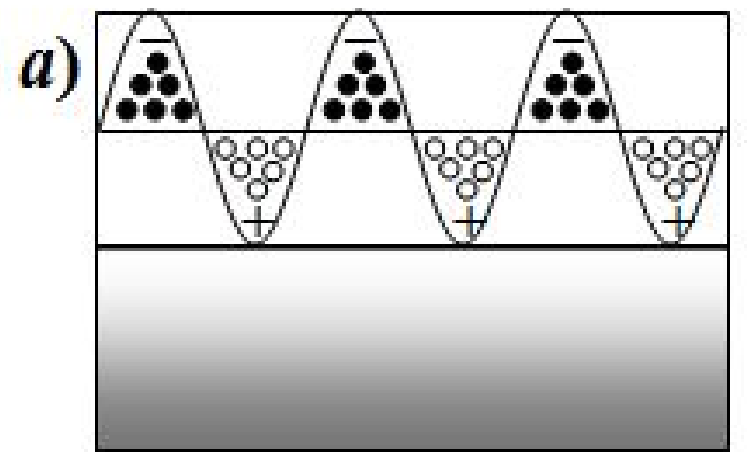}
		\includegraphics[width=50mm]{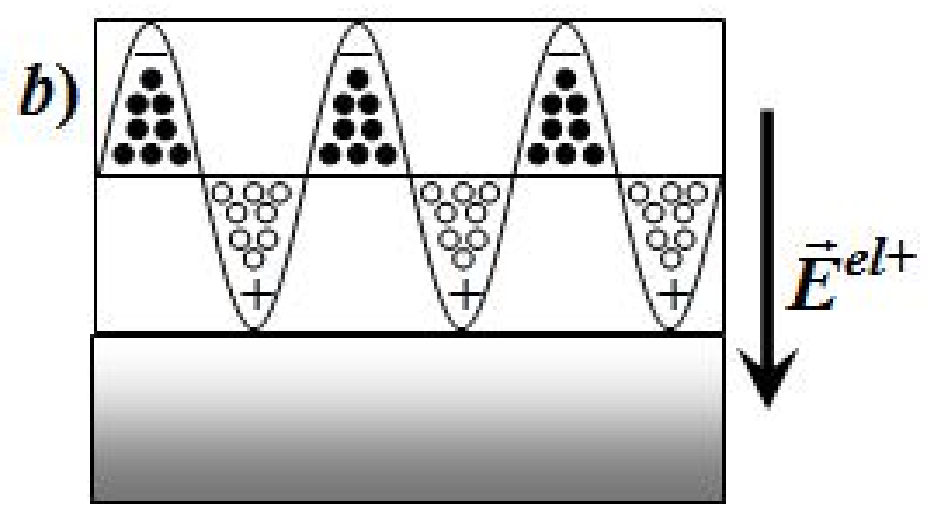}
		\includegraphics[width=50mm]{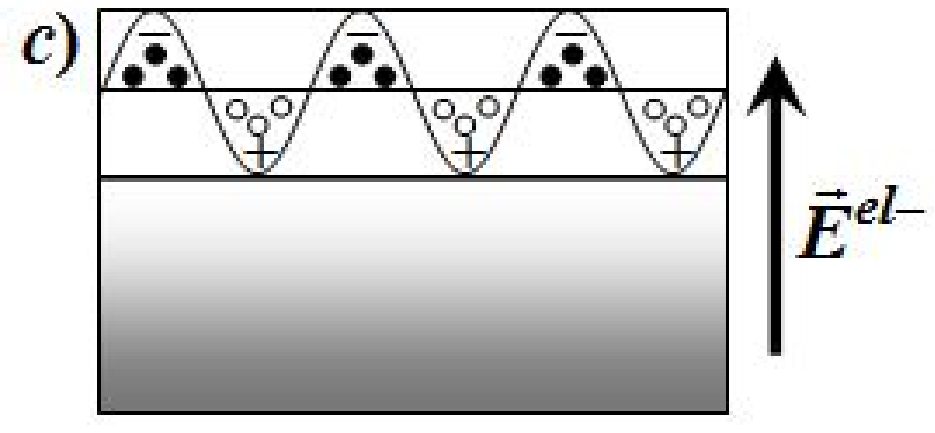}
	\end{multicols}
	\caption{The geometric model of formation of the surface superlattice of adatoms under the influence of the electric field. Here, black circles are the defects of type of the centers of stretching; white circles are the defects of type of the compression centers.}
	\label{fig1}
\end{figure}

Thus, on the surface of semiconductor, there is not only a periodic modulation of the surface relief with the accumulation of adatoms in the maxima or minima of deformation (depending on the sign of the deformation potential $\theta_\text{d}$), but also the surface modulation of the electronic density [figure~\ref{fig1}~(a)]. As can be seen from the formulae (\ref{eq4}), (\ref{eq6}),  (\ref{eq7}),  (\ref{eq16}) and (\ref{eq17}), the periodically distributed deformation $\varepsilon(x)$, the surface concentration of adatoms $N_1(x)$ and the concentration of electrons $n_1(x)$ are in the same phase at $\theta_\text{d} > 0$  (the condition $ql_\text{d} < 1$ is fulfilled) and $a_\text{c} < 0$. For $\theta_\text{d} < 0$, the distribution of the surface concentration of the adatoms $N_1(x)$ and the concentration of electrons $n_1(x)$ are in the opposite phase. Moreover, since for GaAs, the constant of the hydrostatic deformation potential of the conduction band $a_\text{c} < 0$, an excess of electrons will be observed in those areas of the surface where adatoms are accumulated, which are the centers of stretching ($\theta_\text{d} = K\Delta\Omega > 0$, $\Delta\Omega$ is the change in the volume of the crystal by one adsorbed atom, $K$ is comprehensive compression module), and conversely, in those areas of the surface where adatoms  are accumulated, which are the centers of compression, there will be a shortage of electrons in comparison with the mean value [figure~\ref{fig1}~(a)]. That is, the defects of the type of the centers of stretching and the electrons accumulate in the deformation maxima. On the contrary, the defects of the type of the centers of compression accumulate in the deformation minima and there is a decrease in the concentration of electrons in comparison with their spatially uniform value. This is explained by the fact that the deformation flow of electrons ($j_\text{el} = - a_\text{c} n\mu _n \frac{{\partial \varepsilon }}{{\partial x}}$) is directed toward an increase of deformation and the electrons are localized in the region with a lower potential energy (there is a local shift of the bottom of the conduction band by the value of $\Delta E_\text{c}=a_\text{c}\varepsilon$ ). At the same time, the deformation flow of adatoms [the second term in the formula (\ref{eq10})], which are the centers of stretching, is also directed toward an increase of deformation, and the centers of compression are directed towards the reduction of deformation. The controlled parameter $N_0$  is determined by the intensity of laser irradiation.

By placing the semiconductor in the electric field, perpendicular to the direction of propagation of the acoustic wave [figure~\ref{fig1}~(b), (c)], an additional pressure on the surface of semiconductor is created:
\begin{equation}
\label{eq18}
\sigma _{zz}^\text{el}  = e n\left( x \right)E^\text{el}.
\end{equation}

Moreover, in the case where the vector of the electric field strength $\vec E^\text{el}$ is directed into the crystal depth [figure~\ref{fig1}~(b)], the surface area where the adatoms accumulate, which are the centers of stretching, and, respectively, electrons accumulate, is exposed to additional stretching. The surface area in which the adatoms of type of the compression centers accumulate and there is a shortage of electrons, is exposed to additional compression. This, in turn, leads to the emergence of additional deformation-diffusion flow of adatoms (for defects of type of the centers of stretching in the direction of an increasing deformation, for defects of type of the compression centers --- in the opposite direction). At change of the direction of the electric field [figure~\ref{fig1}~(c)], an additional pressure caused by the external electric field leads to a decrease in the deformation gradients on the surface of  a semiconductor and, accordingly, to the delocalization of adatoms.

The spatial nonuniform distribution of adatoms modulates the surface energy $F(x)$, which leads to the appearance of lateral mechanical tension  $\sigma _{xz}  = \frac{{\partial F(N(x))}}{{\partial x}}$, which is compensated by a shift tension in the medium~\cite{Eme08}: 

$
F(N(x)) \approx F(N_0 ) + \left. {\frac{{\partial F}}{{\partial N}}} \right|_{N = N_0 } N_1 (x)$,  $\left. {\frac{{\partial F}}{{\partial N}}} \right|_{N = N_0 }  \approx \frac{{\theta_\text{d}^2 N_0 }}{{aK}}$~\cite{Pel14}.

The boundary condition expressing the balance of lateral tensions is as follows:
\begin{equation}
\label{eq19}
\frac{E}{{2\left( {1 + \nu } \right)}}\left( {\frac{{\partial u_x }}{{\partial z}} + \frac{{\partial u_z }}{{\partial x}}} \right)_{z = 0}  = \frac{{\partial F(N(x))}}{{\partial x}} = \frac{{\partial F}}{{\partial N}}\frac{{\partial N_1 (x)}}{{\partial x}}\,,
\end{equation}
where $E$ and $\nu$ are Young's modulus and Poisson's ratio, respectively.

Besides, the interaction of adatoms with atoms of a semiconductor results in the emergence of the normal mechanical tension on the surface, and the corresponding boundary condition is of the form:
\begin{equation}
\label{eq20}
\frac{{E\left( {1 - \nu } \right)}}{{\left( {1 + \nu } \right)\left( {1 - 2\nu } \right)}}\left( {\frac{{\partial u_z }}{{\partial z}} + \frac{\nu }{{1 - \nu }}\frac{{\partial u_x }}{{\partial x}}} \right)_{z = 0}  = \frac{{\theta_\text{d} }}{a}N_1 (x) + e n\left( x \right)E.
\end{equation}

Thus, a system of homogeneous linear equations (\ref{eq19}) and (\ref{eq20}) for amplitudes $R$ and $Q$ is obtained and the dispersion dependences  $\lambda(q)$ can be obtained from the condition of non-triviality of solutions (from the condition of equality to zero of the determinant of this system).

\section{Calculation results and their discussion}

The calculations of $\lambda(q)$ were carried out for GaAs semiconductor at the following values of pa\-ra\-me\-ters: $N_0 = 2\cdot10^{12}~\text{cm}^{-2}$; $l_\text{d} = 2.9~\text{nm}$; $a = 0.565~\text{nm}$; $c_\text{l} = 3500~\text{m/s}$;  $c_\text{t} = 2475~\text{m/s}$;  $\rho = 5320~\text{kg/m}^3$;  $a_\text{c} = -7.17~\text{eV}$;  $\theta_\text{d} = 10~\text{eV}$;  $D_\text{d} = 10^7 \exp \left( { - \frac{{5.6~\text{eV}}}{{k_\text{B} T}}} \right)~\text{cm}^{2}/\text{s}$~\cite{Wag91};  $\tilde \varepsilon = 12$. Mobility of electrons as function of temperature and concentration was determined by the technique given in work~\cite{Mna04}.

\begin{figure}[!t]
  	\vspace{7mm}
      \begin{multicols}{2}
                    \includegraphics[width=77mm]{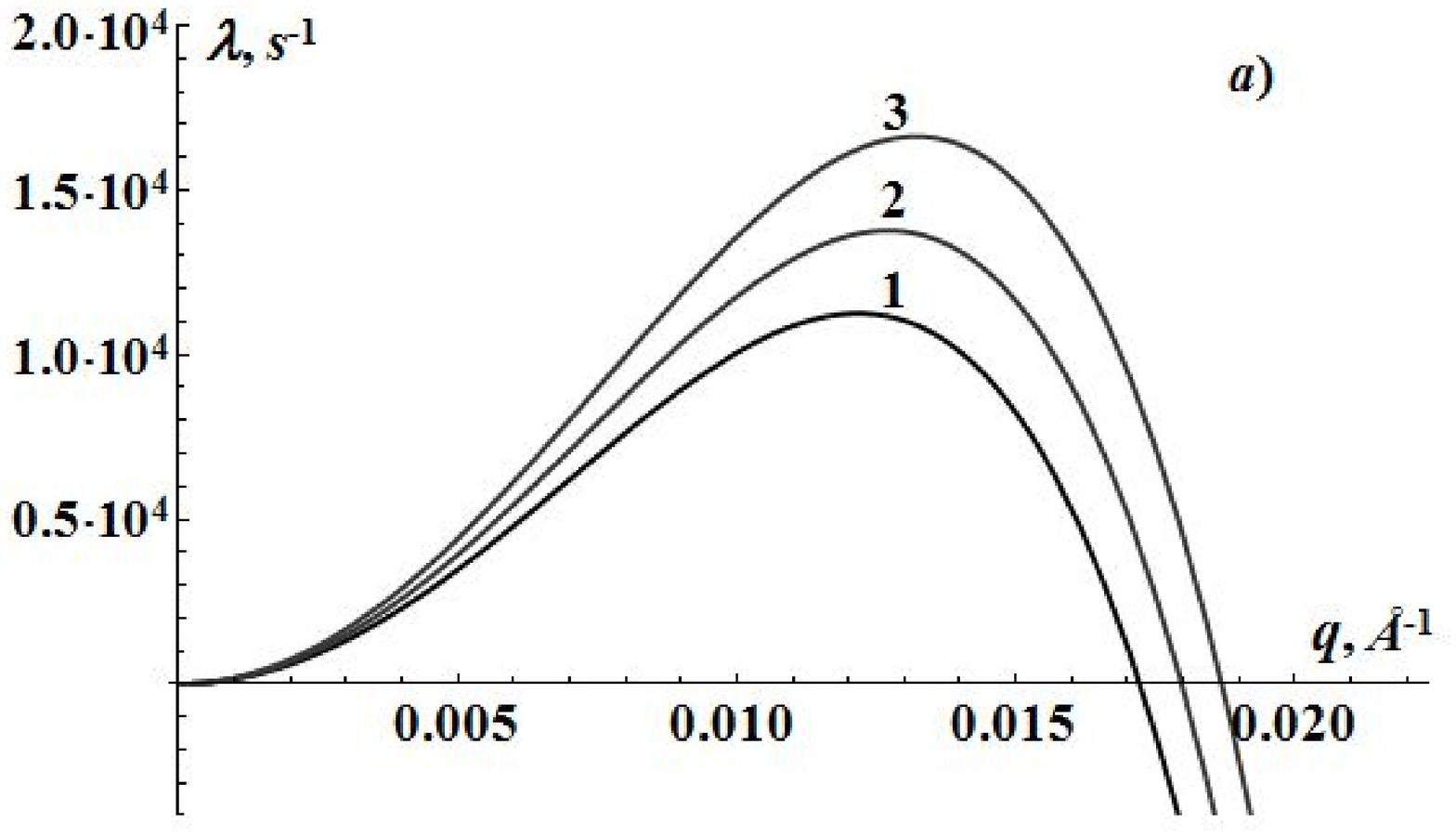}
                    \includegraphics[width=77mm]{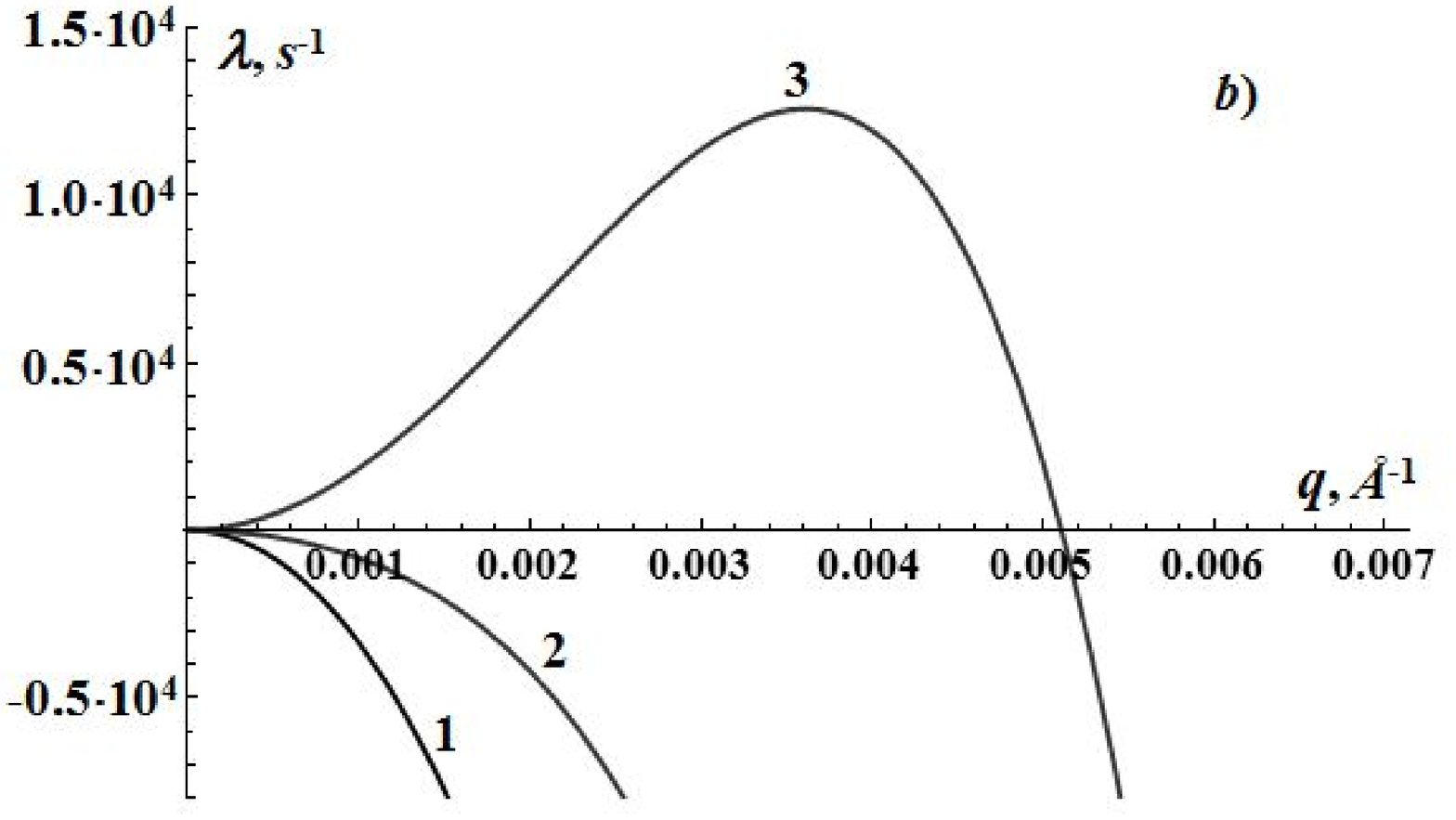}
                    \includegraphics[width=77mm]{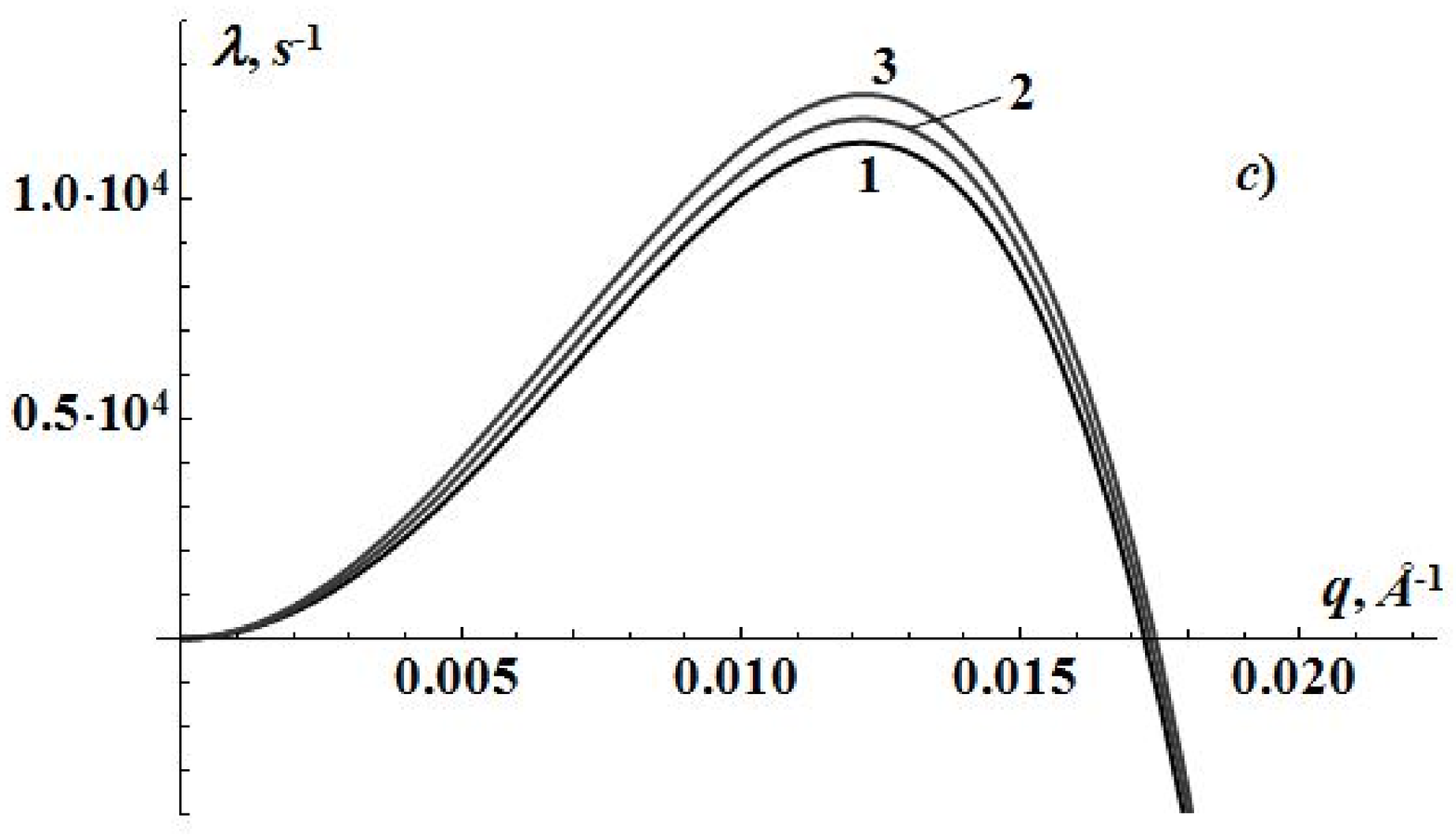}
        \caption{Dependence of the increment of defect-deformation instability on the wave vector at the following values of parameters:
$T=70$~K [(a), (c)];  $T=100$~K~(b);  $n_0=10^{12}~\text{cm}^{-2}$ [(a), (b)];  $n_0=10^9~\text{cm}^{-2}$~(c);
$1-E^\text{el}=0$; $2-E^\text{el+} = 30~\text{kV/cm}$; $3-E^\text{el+}=80~\text{kV/cm}$.}
        \label{fig2}
                    \includegraphics[width=70mm]{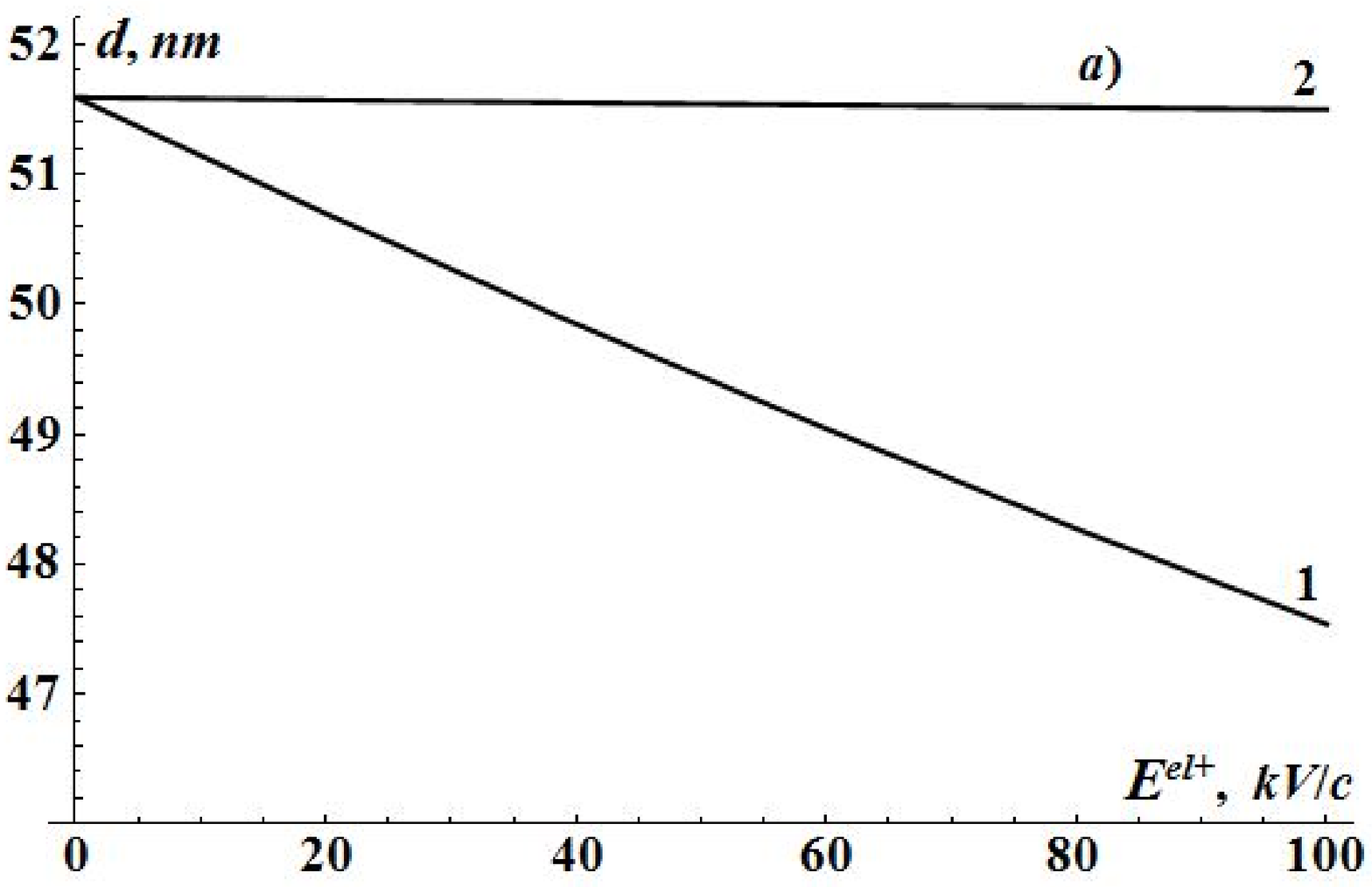}
                    \includegraphics[width=70mm]{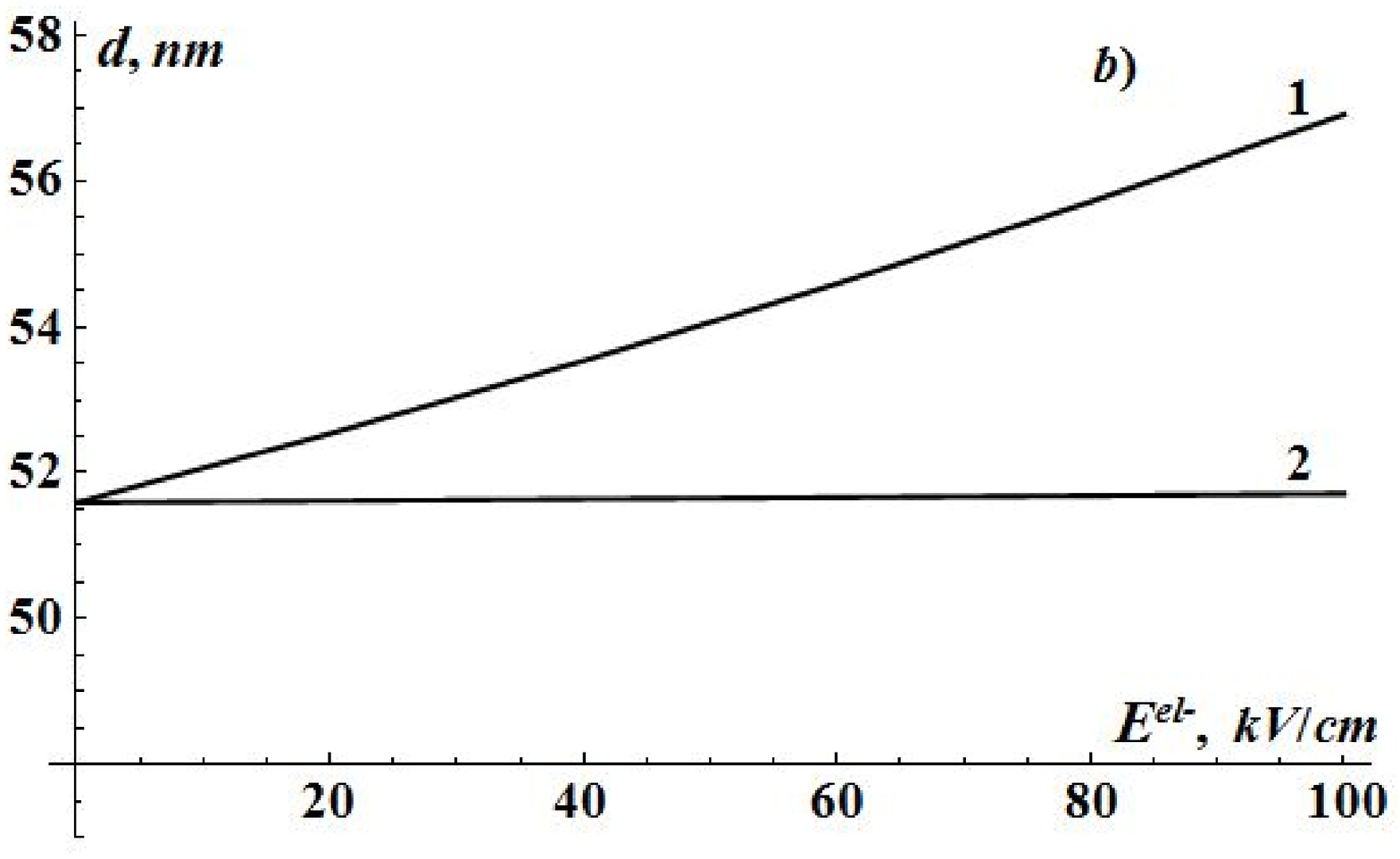}
                    \includegraphics[width=70mm]{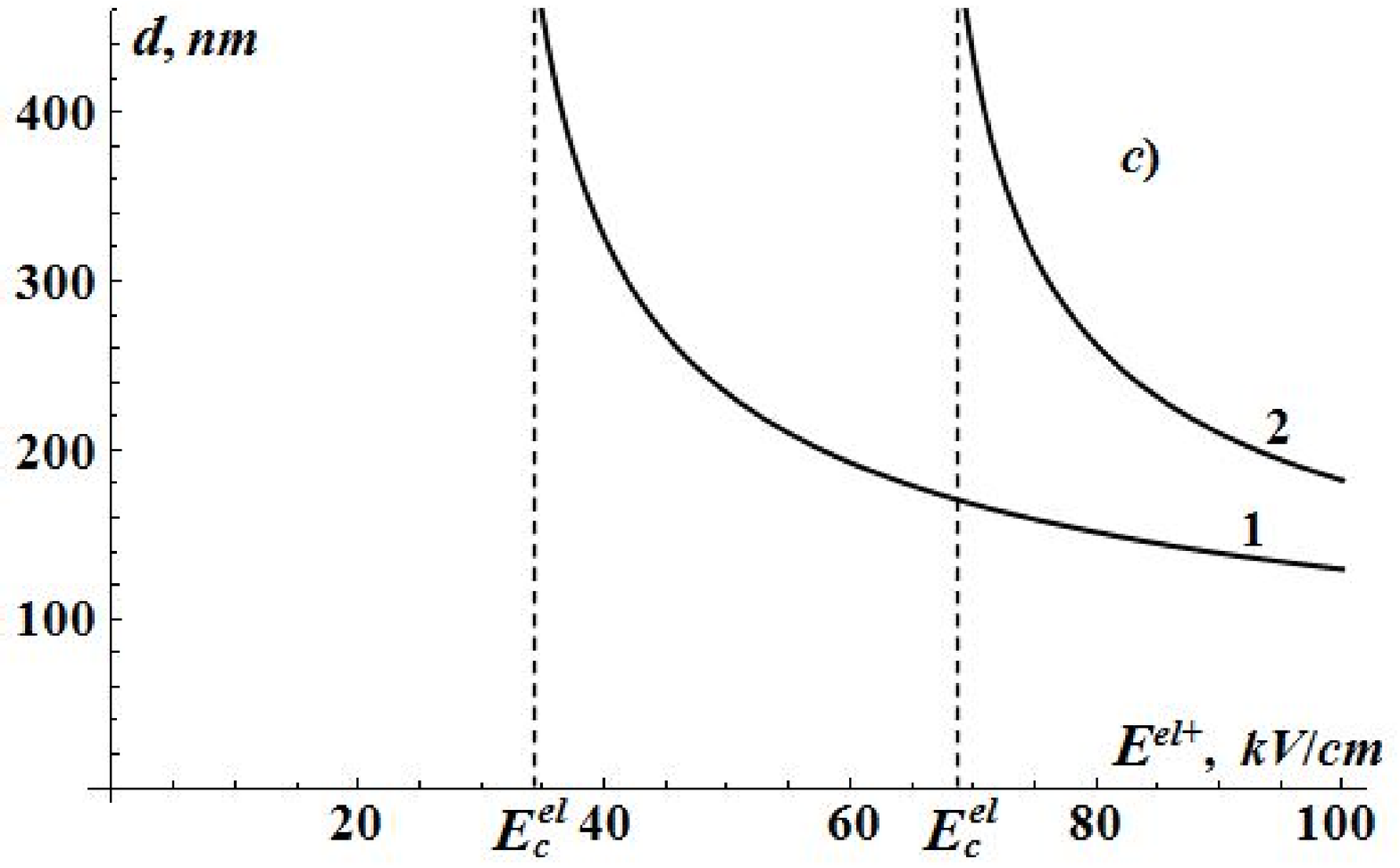}
        \caption{Dependence of the period of the surface defect-deformation structure  on the electric field strength at its various directions
 and at various va\-lu\-es of the electron concentration:
 $T=70$~K~[(a), (b)];  $T=100$~K~(c); $1-n_0=10^{12}~\text{cm}^{-2}$;  $2-n_0=10^9~\text{cm}^{-2}$.}
        \label{fig3}
    \end{multicols}
    \end{figure}

Figure~\ref{fig2} shows the results of calculation of the dependence of the increment of defect-deformation instability on the module of the wave vector at various values of the concentration of conduction electrons, temperature and electric field strength ($E^\text{el+}$) with the direction shown in figure~\ref{fig1}~(b). Such a dependence has a maximum, which is shifted towards great values of the module of the wave vector with an increase in the electric field strength. The formation of the surface superlattice of adatoms is possible only at positive values of the increment of defect-deformation instability  $\lambda$. As seen from figure~\ref{fig2}, the formation of the superlattice is possible only at a temperature lower than a certain critical value $T_\text{c}$. The formation of the surface superlattice is defined by the ratio between the ordinary diffusion flow of adatoms [the first term of equation (\ref{eq10})] and the deformation flow [the second term of equation  (\ref{eq10})]. At high temperatures, the first term is decisive and over time the concentration of adatoms on the surface is levelled, and the formation of the surface superlattice is impossible ($\lambda < 0$). As the temperature decreases, the contribution of the deformation flow of adatoms increases, which becomes decisive at values of temperature less than the critical value of $T_\text{c}$. In this case, the defects accumulate in the maxima (minima) of deformation and the surface superlattice is formed. 
In particular, at a temperature of 100~K [figure~\ref{fig2}~(b)], in the absence of the electric field, the formation of the surface superlattice at a given intensity of laser irradiation is impossible. However, an increase in the electric field strength leads to a change in the sign of the increment of defect-deformation instability $\lambda$  [figure~\ref{fig2}~(b), curve 3], which makes the formation of the superlattice of adatoms possible. Reducing the electron concentration at a constant temperature leads to a decrease in the value of the increment of defect-deformation instability [figure~\ref{fig1}~(c)]. Thus, in a semiconductor with a higher degree of doping with donor impurities, the processes of the formation of nanometer periodic structures should be faster.

\begin{figure}[!t]
	\begin{center}
                          \includegraphics[width=95mm]{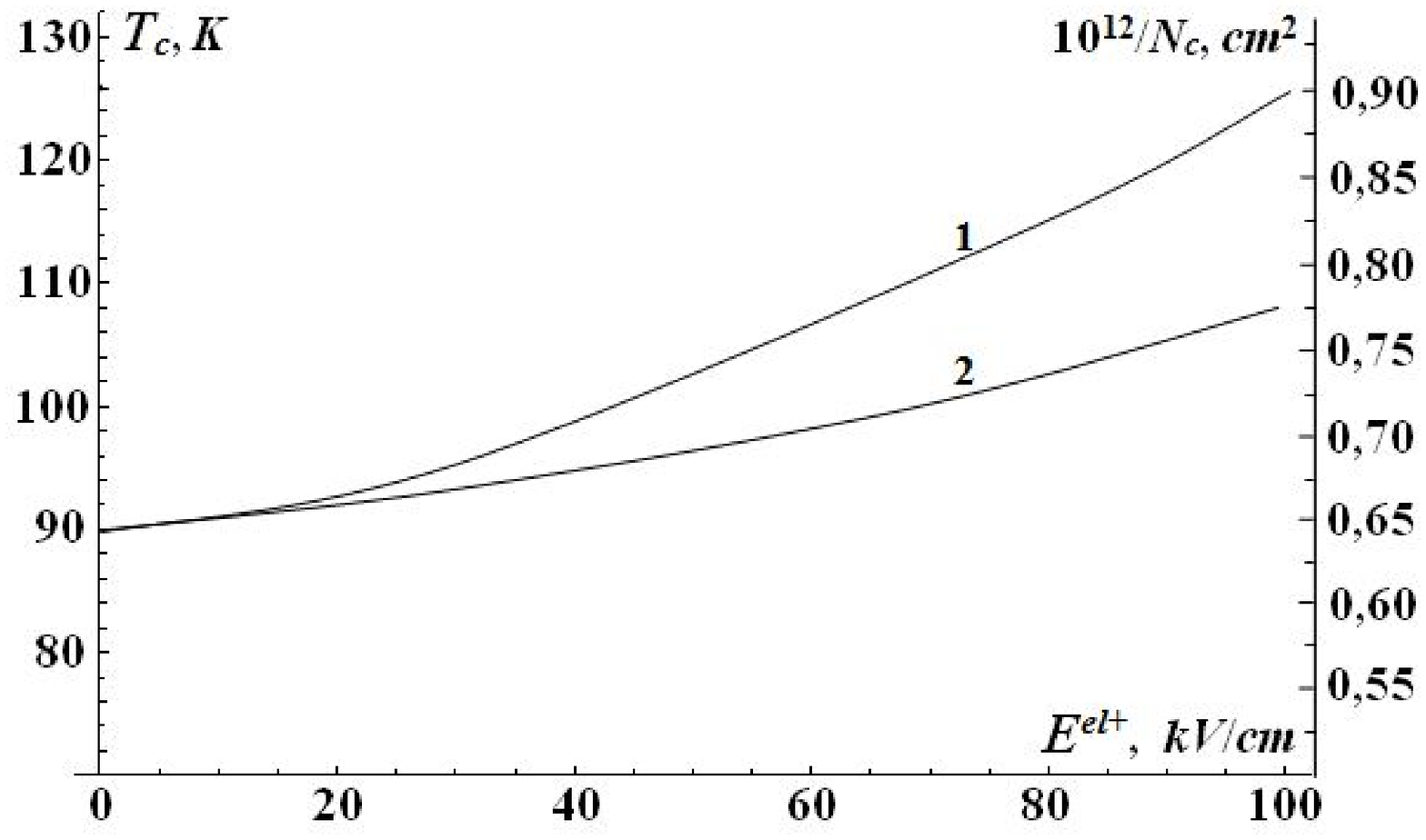}
        \caption{Dependence of the critical temperature (critical concentration) on the electric field strength at different values of the electron concentration:
$1-n_0=10^{12}~\text{cm}^{-2}$;  $2-n_0=10^9~\text{cm}^{-2}$.}
        \label{fig4}
        \end{center}
    \end{figure}

The value of $q_\text{max}$, at which the increment of defect-deformation instability has a maximum, determines the period of the dominant structure $d = \frac{{2\piup }}{{q_{\max } }}$  (figure~\ref{fig3}). Figure~\ref{fig3} shows the dependence of the period of the surface superlattice of adatoms on the electric field strength at various values of the electron concentration, temperature and various directions of the electric field. An increase in the electric field strength leads to a decrease [figure~\ref{fig3}~(a),~(c)] and increase [figure~\ref{fig3}~(b)] of the period of the surface superlattice of adatoms, depending on the direction of the electric field. The effect of the electric field on the period of the surface defect-deformation structure manifests only in strongly alloyed semiconductors. In particular, in GaAs semiconductor with the surface electron concentration of $n_0 = 10^{12}~\text{cm}^{-2}$, at a temperature of 70~K, when the electric field strength increases to $60~\text{kV/cm}$, the period of the superlattice changes by $2.5~\text{nm}$. In a semiconductor with the surface electron concentration of $n_0 = 10^{9}~\text{cm}^{-2}$, the period of the superlattice practically does not change~[figure~\ref{fig3}~(a),~(b)].

At temperatures that are insignificantly higher than the critical temperature $T_\text{c}$ [figure~\ref{fig3}~(c)], there is a significant effect of the electric field, and the direction of the electric field ($E^\text{el+}$) corresponds to~figure~\ref{fig1}~(b). In this case, in the absence of the electric field, the defect-deformation structures do not arise at the given intensity of laser irradiation. However, there is some critical value of the electric field strength  $E_\text{c}^\text{el}$, at the excess of which the formation of the surface periodic structure of adatoms is possible. Otherwise, it can be interpreted as expansion of temperature intervals under the action of the electric field, within which the formation of the surface superlattice is possible.

Changing the direction of the electric field [figure~\ref{fig1}~(c)] leads to deterioration of the conditions of the formation of self-organized nanostructures, irrespective of the sign of the deformation potential  $\theta_\text{d}$, in particular, it leads to a decrease in the critical temperature and to an increase of the period of the surface superlattice [figure~\ref{fig3}~(b)]. This is explained by the fact that in this case, the local deformation caused by the action of the electric field bears the character opposite to the deformation created by adatoms~(figure~\ref{fig1}).

The intensity of laser irradiation (the average concentration of adatoms) is another parameter that determines the conditions for the nucleation of the surface superlattice of adatoms. Figure~\ref{fig2} shows the dependence of the increment of defect-deformation instability on the wave vector at different values of temperature and at the fixed value of $N_0 = 2\cdot10^{12}~\text{cm}^{-2}$. Similar dependencies can be obtained at various values of $N_0$, having fixed a certain value of temperature. In this case, there is a critical value of the concentration of adatoms $N_\text{c}$. For concentrations less than $N_\text{c}$, the formation of the periodic surface structure is impossible. Figure~\ref{fig4} shows the dependence of the critical temperature $T_\text{c}$ (the critical value of the concentration of adatoms $1/N_\text{c}$) on the electric field strength at different values of the concentration of electrons. The area under the curves is the area in which there is formed a periodic structure of adatoms.  The electric field, whose direction corresponds to figure~\ref{fig1}~(b), allows one to reduce the critical value of the concentration of adatoms (the intensity of laser irradiation) or to increase the temperature below which the formation of the surface superlattices occurs. The effect of the electric field is more significant for highly doped semiconductors.

In the framework of this model, the initial (linear) stage of the formation of a surface superlattice is considered ($t < 1/\lambda$). Due to an increase of the amplitude of deformation, nonlinear effects become significant, which leads to its saturation (the amplitude no longer increases). In this case, in order to calculate the amplitude in the energy of elastic interaction, it is necessary to take into account the anharmonic terms ~\cite{Pel14}, or to consider the nonlinearity in the boundary conditions. Except the structure dominating in the linear approximation with an increment $\lambda_\text{max}(q_\text{max})$, the whole continuum of structures with $\lambda> 0$  intensifies (figure~\ref{fig2}). At an exit to the stationary regime, an increase of the amplitude of the dominant structure will reach saturation due to elastic nonlinearity, but the amplitudes with a smaller increment will continue to increase ~\cite{Eme99}. Thus, the spectrum of the generated modes can extend. However, it is shown in~\cite{Eme99}  that in solids with defects which are the centers of tension ( $\theta_\text{d} > 0$), only single-mode generation with $q_\text{max}$ is most frequently realized.

\section{Conclusions}

1.	The theory of nucleation of the surface superlattice of adatoms in GaAs semiconductor under the effect of laser irradiation at the action of electric field, directed perpendicular to the direction of propagation of the SAW, is developed. The proposed theory takes into account the interaction of adatoms and conduction electrons with self-consistent SAW. The semiconductor can be located both in the external electric field and internal one, created, for example, by a hetero-borders. The electric field, transverse to the direction of propagation of the SAW, creates an additional non-uniform mechanical tension. Depending on the direction of the electric field, it is possible to increase or decrease the deformation flows of adatoms.

2.  The formation of the superlattice is possible if the average concentration of adatoms exceeds a certain critical value (or the temperature is less than a certain critical value). The concentration of adatoms is defined by the intensity of laser irradiation. It is established that in GaAs semiconductor, an increase of the electric field strength, depending on the direction, leads to an increase or to a decrease of the critical temperature (the critical concentration of adatoms), at which the formation of self-organized nanostructure is possible. It is shown that the effect of the electric field is more significant in highly doped semiconductors as well as in semiconductors with a high value of the constant of hydrostatic deformation potential of the conduction band and electron mobility. GaAs semiconductor is the most optimal for these parameters.

3. It is shown that in strongly alloyed $n$-GaAs semiconductor, an increase of the electric field strength leads to a monotonous change (decrease or increase depending on the direction of the electric field) of the period of self-organized surface nanostructures of adatoms.

4.	It is established that the effect of the electric field on the conditions of formation and on the period of the surface superlattice does not depend on the sign of the deformation potential of adatoms.

%
%

\ukrainianpart


\title{Вплив електричного поля на  нуклеацію нанометрової періодичної структури адатомів у напівпровіднику GaAs
під впливом лазерного опромінення}
\author[Р.М.~Пелещак, О.В.~Кузик, О.О.~Даньків,  С.К.~Губа]{Р.М.~Пелещак\refaddr{label1}, О.В.~Кузик\refaddr{label1}, О.О.~Даньків\refaddr{label1}, С.К.~Губа\refaddr{label2}}
\addresses{
	\addr{label1} Дрогобицький державний педагогічний університет імені Івана Франка,\\ вул. Івана Франка, 24, 82100 Дрогобич, Україна 
	\addr{label2} Національний університет ``Львівська політехніка'', вул. Степана Бандери, 12, 79013 Львів, Україна
}
%
%
%

\makeukrtitle

\begin{abstract}
\tolerance=3000%
У роботі досліджено вплив електричного поля на умови формування та період поверхневої надгратки адсорбованих атомів у напівпровіднику \textit{n}-GaAs. Встановлено, що у напівпровіднику GaAs збільшення напруженості електричного поля залежно від напрямку призводить до збільшення або зменшення критичної температури (критичної концентрації адатомів), при якій можливе формування самоорганізованої наноструктури. Показано, що у сильнолегованому напівпровіднику \textit{n}-GaAs збільшення напруженості електричного поля призводить до монотонної зміни (зменшення чи збільшення залежно від напрямку електричного поля) періоду самоорганізованих поверхневих наноструктур адатомів.
\keywords нуклеація, електричне поле, адатом, поверхнева надгратка, дифузія, деформація

\end{abstract}

\end{document}